\documentclass[manuscript]{aastex}

\usepackage{xcolor}

\fullcollaborationName{The Friends of AASTeX Collaboration}

\begin{document}

\title{Cosmic ray transport near the Sun}


\author{R.D. Strauss\altaffilmark{1}, J.P. van den Berg\altaffilmark{2}}
\affil{Center for Space Research, North-West University, Potchefstroom, 2522, South Africa}

\and
\author{J.S Rankin}
\affil{Department of Astrophysical Sciences, Princeton University, Princeton, NJ 08540, USA}


\altaffiltext{1}{National Institute for Theoretical and Computational Sciences (NITheCS), South Africa} 
\altaffiltext{2}{South African National Space Agency, Hermanus, South Africa}

\begin{abstract}
The strongly diverging magnetic field lines in the very inner heliosphere, through the associated magnetic focusing/mirroring forces, can, potentially, lead to highly anisotropic galactic cosmic ray distributions close to the Sun. Using a simplified analytical approach, validated by numerical simulations, we study the behaviour of the galactic cosmic ray distribution in this newly explored region of the heliosphere and find that significant anisotropies can be expected inside 0.2 au.
\end{abstract}


\section{Introduction}

The {\it Parker Solar Probe } ({\it PSP}) spacecraft is currently exploring and making the first {\it in-situ} galactic cosmic ray (GCR) measurements in the very inner heliosphere. These include particle gradient measurements \citep[][]{rankinetal2021,rankinetal2021b}, potentially down to the Alfv\'en radius. However, the \citet{parker1965} transport equation, used extensively to simulate GCR transport throughout the heliosphere and valid for a nearly isotropic particle distribution, may not be applicable in this region with strongly converging (as seen from Earth) magnetic field lines. In this paper we present a simplified model to study the (potentially anisotropic) transport of GCR in this newly explored region of the heliosphere.

\section{Field-aligned transport}

We consider only transport along the mean magnetic field, $\vec{B} = B \hat{z}$ and neglect adiabatic energy losses, as well as any 2D cross-field transport processes, including drifts and perpendicular diffusion. For the mean (or background) magnetic field we assume a \citet{parker1958} heliospheric magnetic field (HMF) geometry. In the following sections, we study the field-aligned transport of GCRs using models with varying level of complexity.

\subsection{Loss cone distributions}

In the absence of magnetic turbulence the magnetic moment of particles are conserved during propagation and, along with the conservation of kinetic energy, any mirroring/focusing force is accompanied by an interchange between parallel and perpendicular energy: As a particle moves into a region of larger magnetic field strength, its perpendicular speed increases, with the effect that its parallel speed decreases. Ultimately this causes the particle's motion to be reversed and the particle is mirrored. However, not all particles entering a magnetic bottle will be mirrored. A particle starting out in a region with field strength $B_{\mathrm{Earth}}$ with a pitch-angle $\alpha$, and where $\mu = \cos \alpha$, will not be able to penetrate a region of magnetic field strength $B$, if

\begin{equation}
\label{eq:MirrorCondition}
|\mu| > \mu_c = \sqrt{1 - \frac{B_{\mathrm{Earth}}}{B}}.
\end{equation}

\begin{figure}
    \centering
\includegraphics[width=0.49\textwidth]{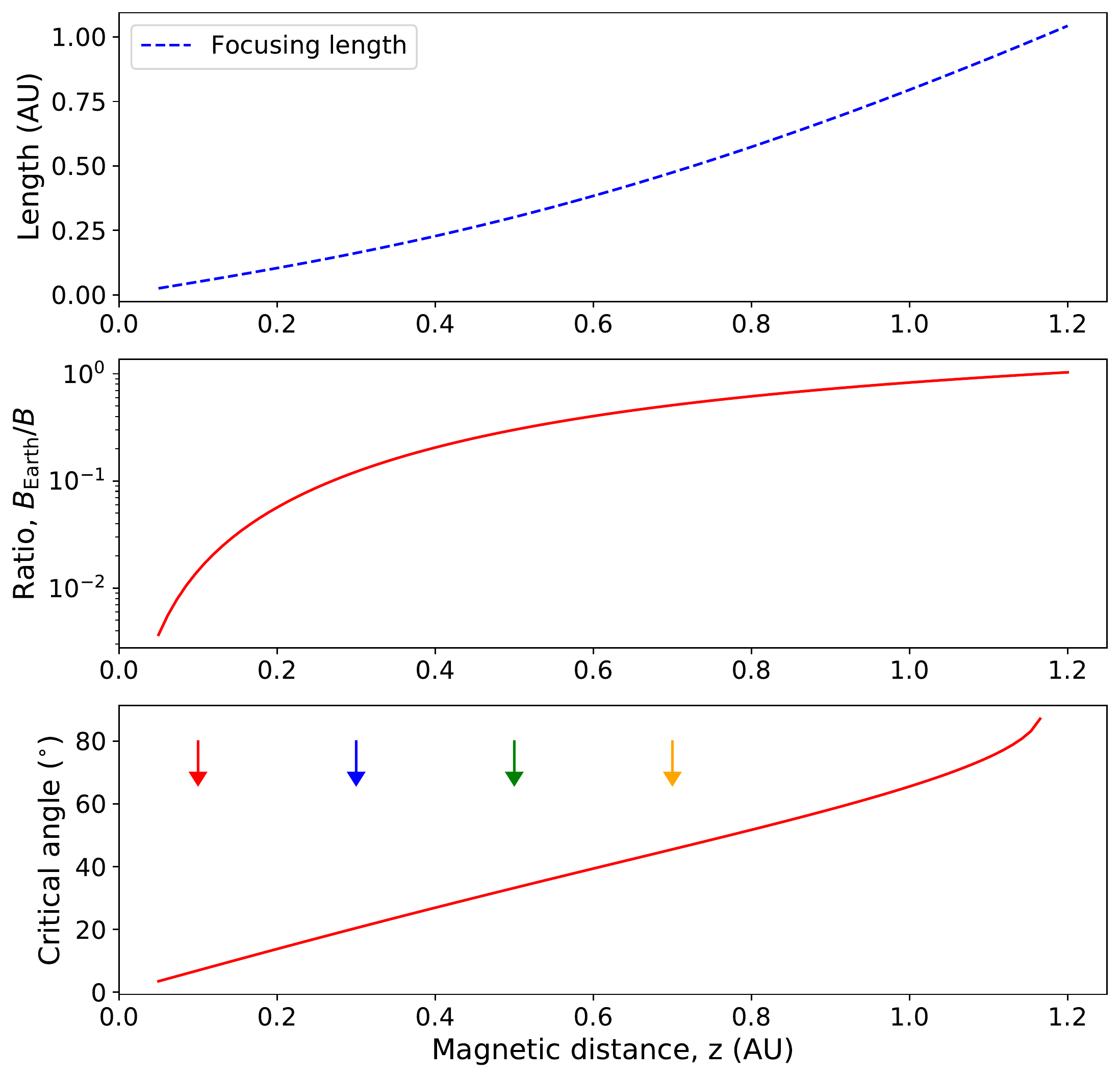}\\
\includegraphics[width=0.23\textwidth]{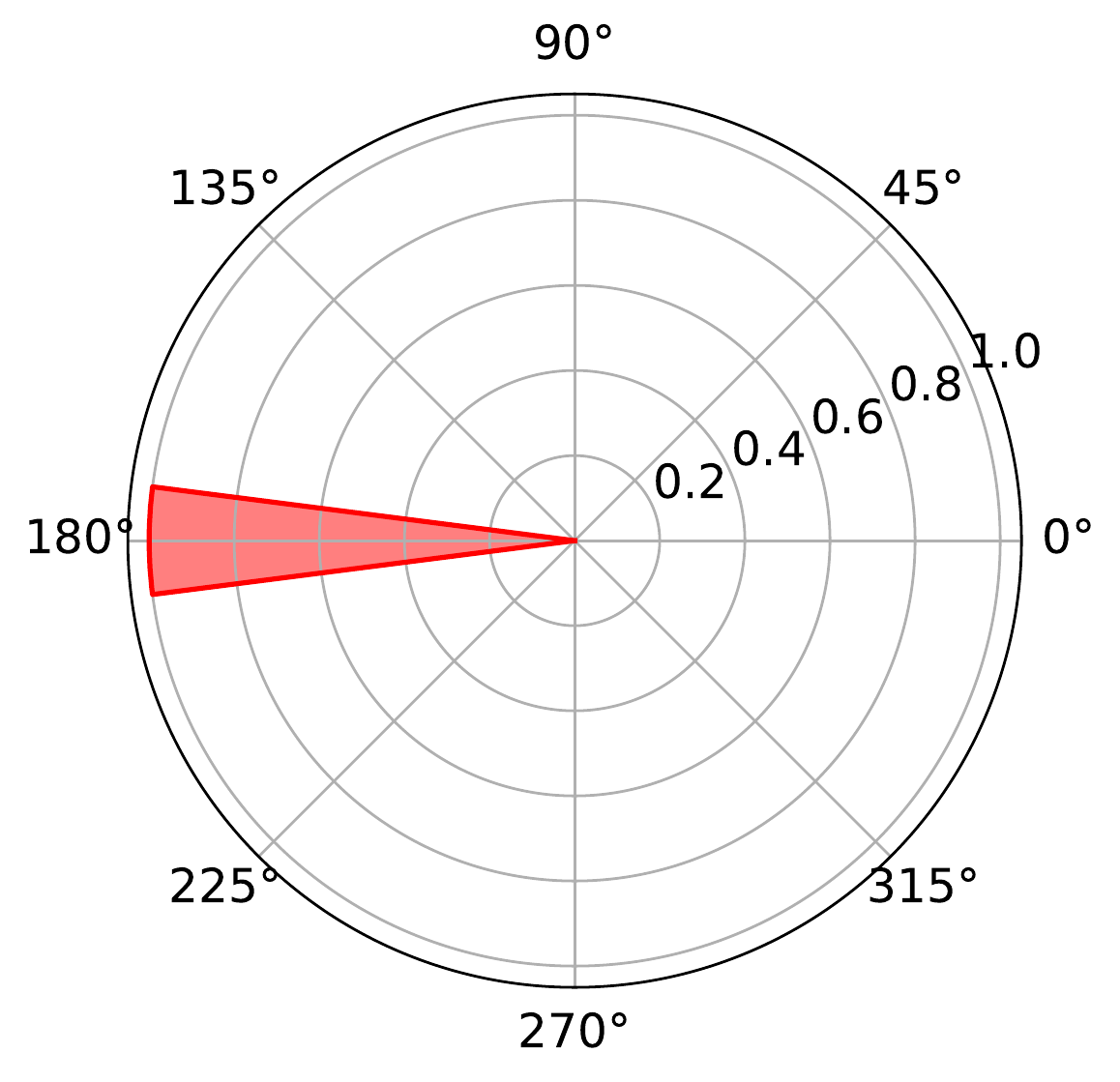}
\includegraphics[width=0.23\textwidth]{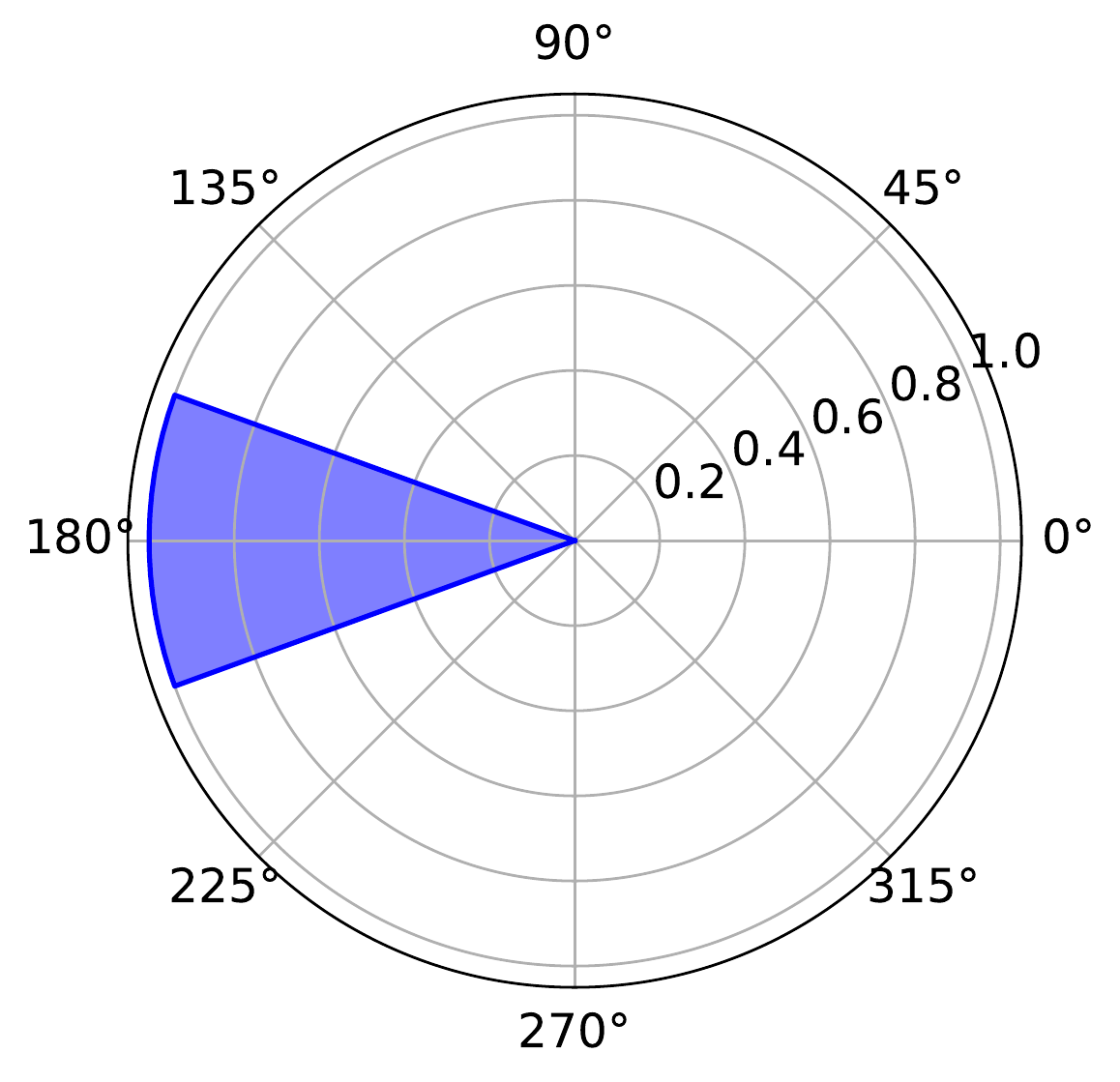}\\
\includegraphics[width=0.23\textwidth]{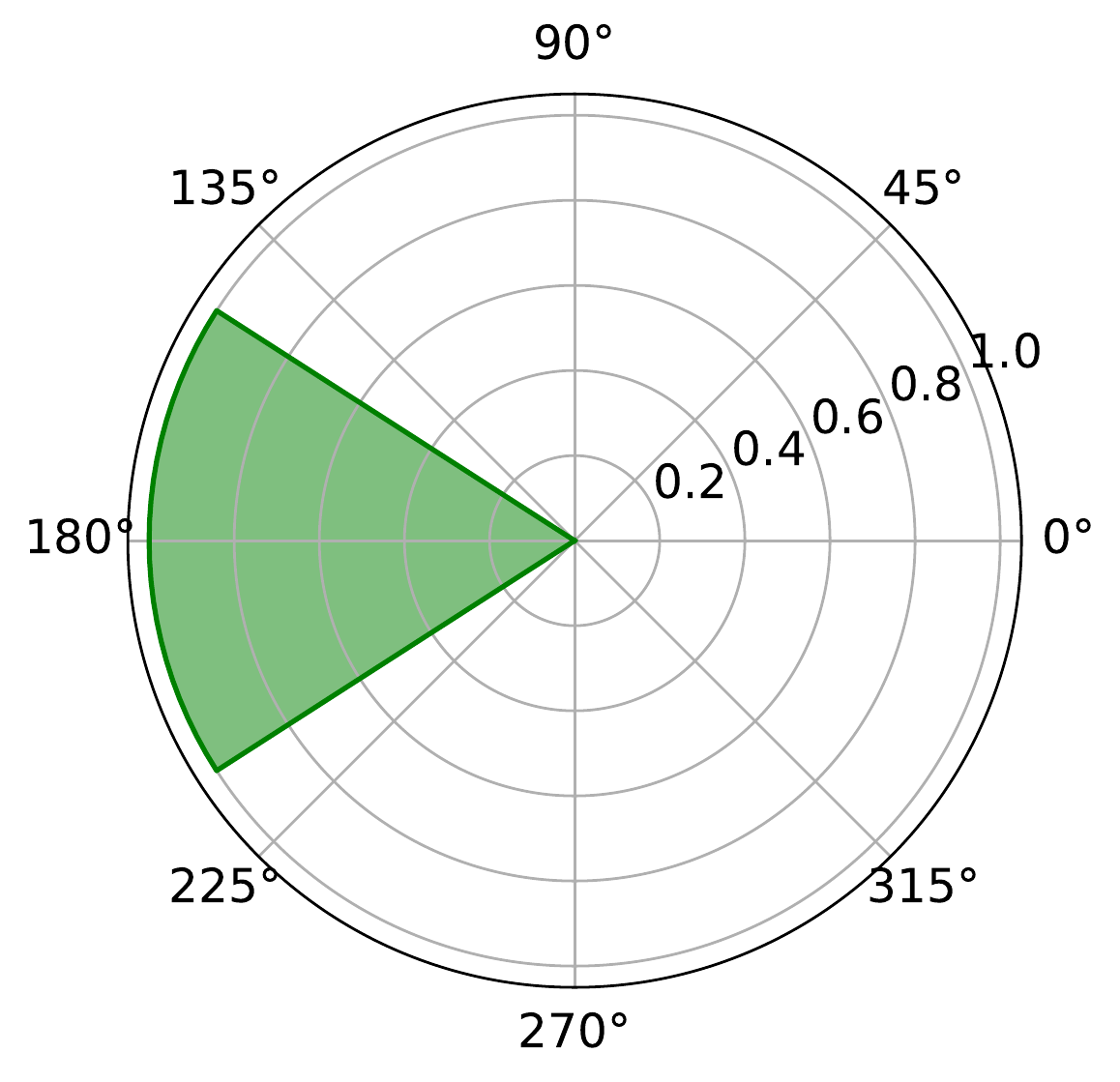}
\includegraphics[width=0.23\textwidth]{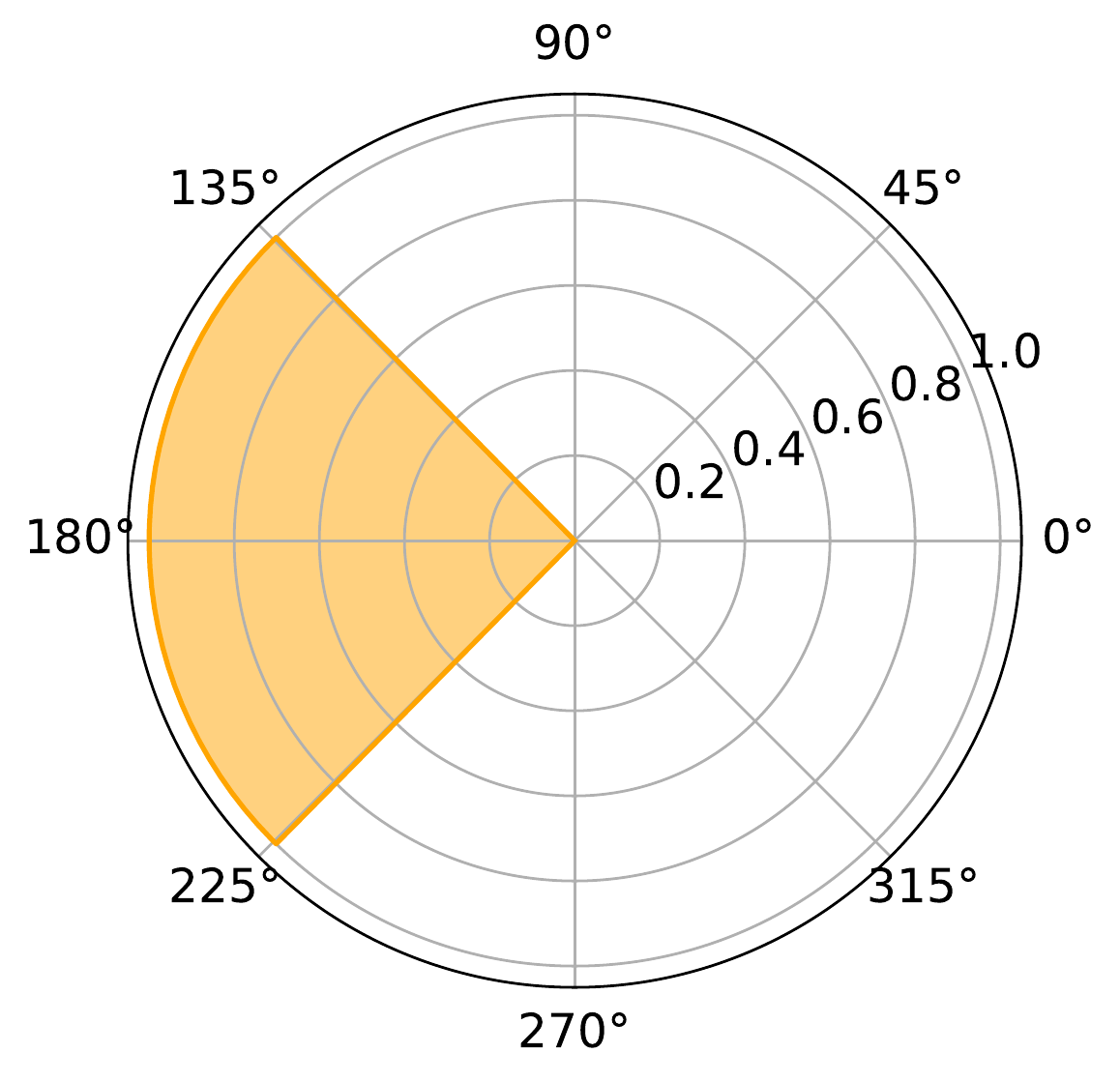}
    \caption{The top panels show the {focusing/mirroring} length, the mirror ratio, and the resulting critical angle, $\alpha_c$, while the bottom panels show the resulting particle distribution at distances of 0.1, 0.3, 0.5, and 0.7 au  (these positions are indicated by the vertical arrows in the upper panels and by colour in the lower diagrams). {Isotropic distributions are assumed at $z=1.2$ au. In the bottom panels, $180^{\circ}$ labels GCRs propagating along the magnetic field towards the Sun and $0^{\circ}$ particles propagating along the magnetic field away from the Sun.}}
    \label{fig:loss_cone}
\end{figure}

We can now calculate this critical angle $\mu_c$ for the case of GCRs, starting out isotropically from Earth, and propagating towards the Sun along $\vec{B}$. In Fig. \ref{fig:loss_cone} we show the mirror ratio $B_{\mathrm{Earth}}/B$, and the resulting $\alpha_c = \cos ^{-1}\mu_c$, as a function of radial distance. {The effectiveness of the focusing/mirroring process decreases quickly away from the Sun and become negligible beyond Earth's orbit as the mirror ratio approaches unity. Without any additional process that can drive anisotropic behaviour, GCR are generally observed to be nearly isotropic after propagating (diffusively) from the heliopause to Earth \citep[see e.g.][]{gilletal2021}.} In the bottom panels of {the figure} we use the calculated $\mu_c$ to estimate the associated {\it loss-cone distributions} at different distances (these distances are indicated by vertical arrows on the upper panels). These loss-cone distribution illustrate how difficult it is for GCR to reach the inner heliosphere; an effect also studied by e.g. \citet{Hutchinsonetal}.

While these results are useful for quantifying the effect of mirroring in the \citet{parker1958} HMF, they do not capture the full picture of particle transport: The HMF is turbulent on all scales and these turbulent fluctuations will lead to pitch-angle scattering that tends to isotropize the GCR distribution and also disrupt the general particle drift picture \citep[e.g.][]{vandenBergetal2020,vandenBergetal2021}. The interplay between scattering (leading to isotropic distributions) and focusing/mirroring (leading to anisotropic distributions) is therefore essential to understand. The additional effect of pitch-angle scattering is considered in the next sections.

\subsection{Focused transport}

The evolution of a nearly gyrotropic distribution function, $f(z, \mu, t)$, is given by the so-called {\it focused transport} equation,

\begin{equation}
\label{Eq:focused_eq}
    \frac{\partial f}{\partial t} + \mu v \frac{\partial f}{\partial z}  +  \frac{v}{2L} \left( 1-\mu^2 \right) \frac{\partial f}{\partial \mu} =   
 \frac{\partial}{\partial \mu} \left(D_{\mu\mu}  \frac{\partial f}{\partial \mu} \right)
\end{equation}

where $\mu$ is the cosine of the particle pitch-angle, $v$ is the particle speed so that $v_{||} = \mu v$ is the parallel (with respect to the magnetic field) projection of the particle speed. Particle mirroring/focused is included via the {focusing/mirroring} length

\begin{equation}
\label{Eq:def_foc_length}
    \frac{1}{L} = -\frac{1}{B} \frac{dB}{dz}.
\end{equation}

and pitch-angle diffusion is included via the pitch-angle diffusion coefficient $D_{\mu \mu}$. Eq. \ref{Eq:focused_eq} is also referred to as the \citet{roelofeq} equation.

Once $f$ is obtained, an omni-directional distribution function can be calculated as

\begin{equation}
\label{Eq:omni_directional}
    F(z,t) = \frac{\int_{-1}^{+1} f(z,\mu,t) d\mu}{ \int_{-1}^{+1} d\mu} =  \frac{1}{2}\int_{-1}^{+1} f(z,\mu,t) d\mu ,
\end{equation}

with an associated first-order anisotropy, given by

\begin{equation}
    A(z,t) = 3 \frac{\int_{-1}^{+1} \mu f(z,\mu,t) d\mu}{\int_{-1}^{+1} f(z,\mu,t) d\mu} = \frac{3}{2F(z,t)} \int_{-1}^{+1} \mu f(z,\mu,t) d\mu,
\end{equation}

and varying between $-3$ (all particles propagating towards the Sun) and $+3$ (all particles propagating away from the Sun).

The efficiency of pitch-angle scattering is usually quantified by the parallel mean free path, calculated here following \citet{HasseL_wibb} as

\begin{equation}
\label{Eq:lamda_def}
    \lambda^0_{\parallel} = \frac{3}{8} v \int_{-1}^1 \frac{(1 - \mu^2)^2}{D_{\mu \mu} (\mu)} {\rm d}\mu,
\end{equation}

which is related to the parallel diffusion coefficient through

\begin{equation}
\label{Eq:kappa_paral}
    \kappa^0_{||} = \frac{v}{3} \lambda^0_{||}.
\end{equation}

For this work, we assume $\lambda^0_{\parallel} = \lambda$ to be a constant.

\subsection{Nearly isotropic formulation with weak focusing}

Following \citet{litvinenkoschlikkie2013}, we can derive an approximate transport equation for $F(z,t)$ from Eq. \ref{Eq:focused_eq} by expanding $f$ in terms of $F$ and a small correction factor $F_1(z,\mu,t)$,

\begin{equation}
    f(z, \mu, t) = F(z,t) + F_1 (z, \mu, t)
\end{equation}

where

\begin{equation}
    \int  F_1 d\mu = 0
\end{equation}

and $\left | F_1 \right| \ll f$. This corresponds to the case of a {\it nearly isotropic} distribution, which is possible when $\lambda^0_{||} \ll L$ so that the transport is diffusion dominated in this {\it weak {focusing/mirroring}} limit. Note that, from here, the explicit dependence on the quantities $z$, $\mu$, and $t$ are not included. By furthermore assuming that $\lambda^0_{||}$ and $L$ are constant, \citet{litvinenkoschlikkie2013} show that the evolution of $F$ is governed by the following pseudo-diffusion equation

\begin{equation}
\label{Eq:pseudo_diff_eq}
\frac{\partial F}{\partial t} + \frac{\kappa_{||}}{L} \frac{\partial F}{\partial z} = \kappa_{||} \frac{\partial^2 F}{\partial z^2},
\end{equation}

which is in the form of a diffusion equation, but with the addition of a coherent advection speed 

\begin{equation}
    u_{||} = \frac{\kappa_{||}}{L}.
\end{equation}

Note that the $\kappa_{||}$ introduced here differs from $\kappa^0_{||}$ as defined in Eq. \ref{Eq:kappa_paral} and includes a correction to account for {focusing/mirroring} effects.  However, in the weak {focusing/mirroring} case considered here we can approximate $\kappa_{||} \approx \kappa^0_{||}$ and use the two quantities interchangeably.

The diffusive streaming flux for this scenario is given by 

\begin{equation}
\label{Eq:streaming_flux}
    S = \frac{v}{2} \int \mu F_1 d\mu \stackrel{\lambda_{||} \ll L}{=} - \kappa_{||} \frac{\partial F}{\partial z}.
\end{equation}

In deriving the results presented here, \citet{litvinenkoschlikkie2013} assumes a constant $L$, which is definitely not the case for a \citet{parker1958} HMF; see the top panel of Fig. \ref{fig:loss_cone}. However, these authors also state that the analytical approximation should still be valid as long as $L$ is constant on the scale of $\lambda_{||}$. This is a reasonable approximation for the constant (i.e. independent of radial distance) $\lambda_{||}$, with values between 0.1 -- 1 au, assumed in this work. However, modeling results of mostly solar energetic particles have shown that $\lambda_{||}$ can vary significantly from event-to-event \citep[e.g.][]{droge2000}, while theoretical calculations indicate that $\lambda_{||}$ can have a complicated radial dependence inside 1 au \citep[e.g.][]{straussleroux2019}.\\
\subsection{Isotropic formulation with no focusing}

In the case of vanishing {focusing/mirroring}, $L \rightarrow \infty$, we have the evolution of the isotropic diffusion equation

\begin{equation}
\label{Eq:diff_eq}
\frac{\partial F}{\partial t}  = \kappa_{||} \frac{\partial^2 F}{\partial z^2},
\end{equation}

which follows directly from Eq. \ref{Eq:pseudo_diff_eq}.


\begin{figure*}
    \centering
    \includegraphics[width=0.49\textwidth]{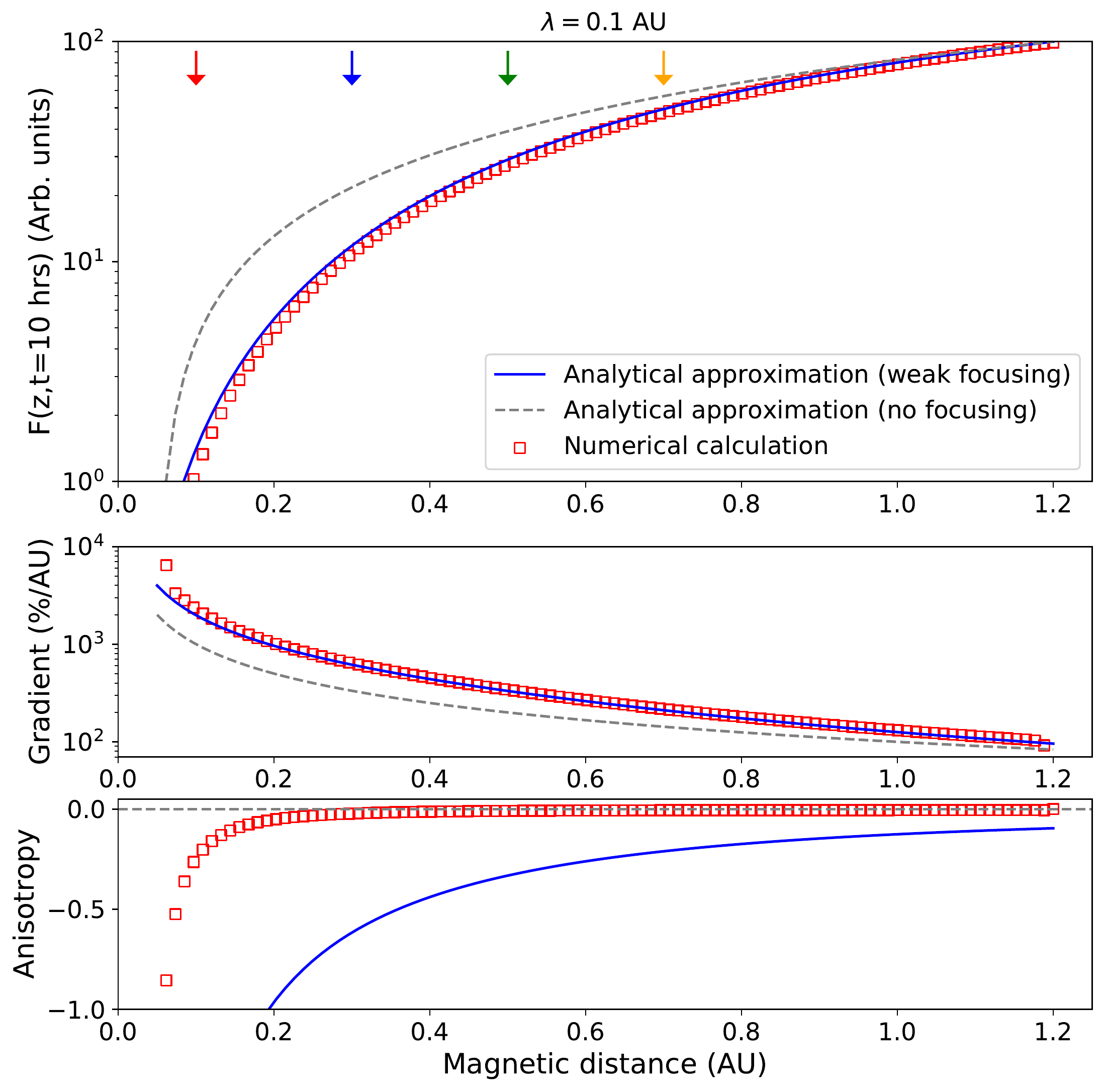}
    \includegraphics[width=0.49\textwidth]{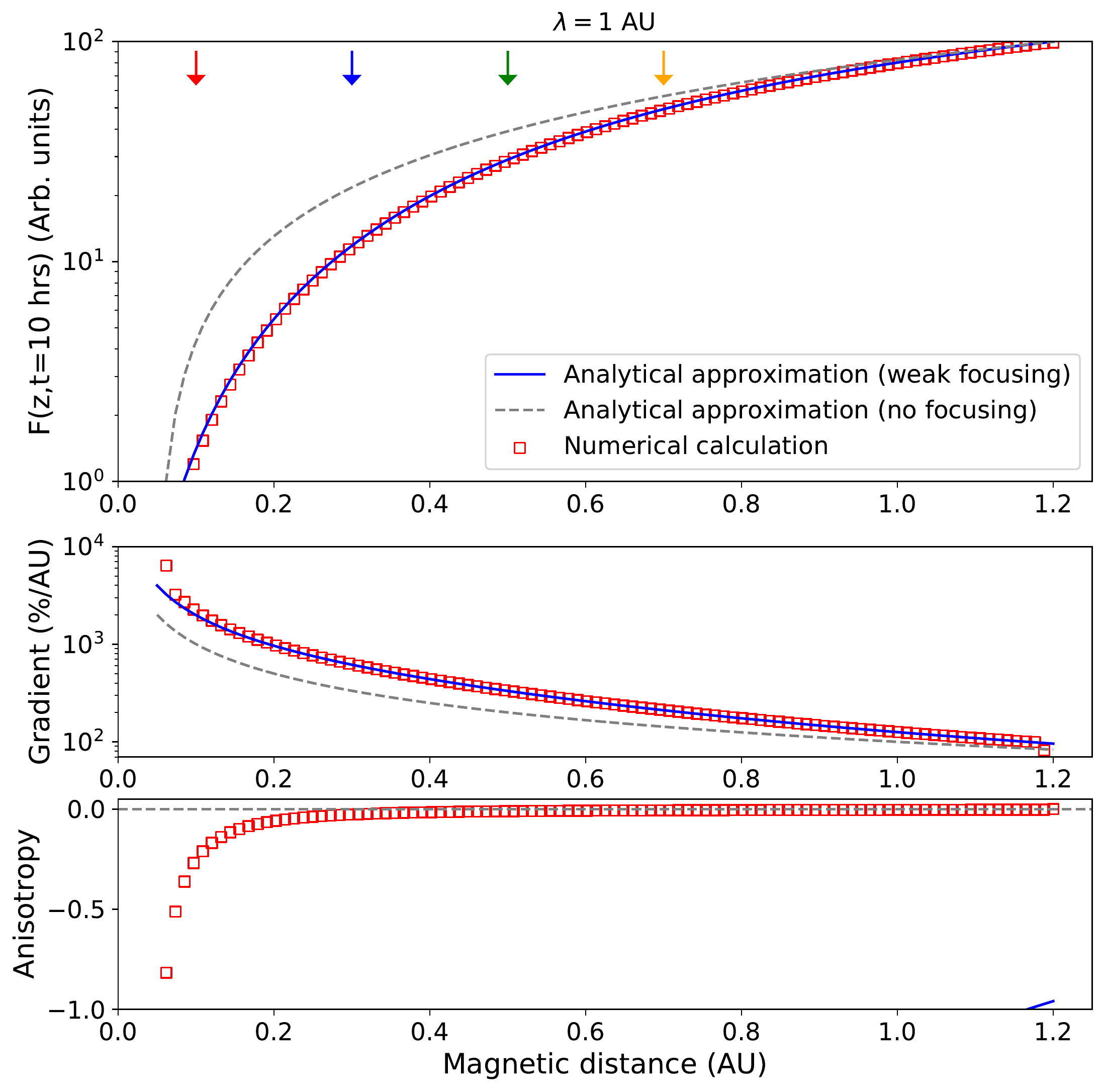}\\
    \includegraphics[width=0.23\textwidth]{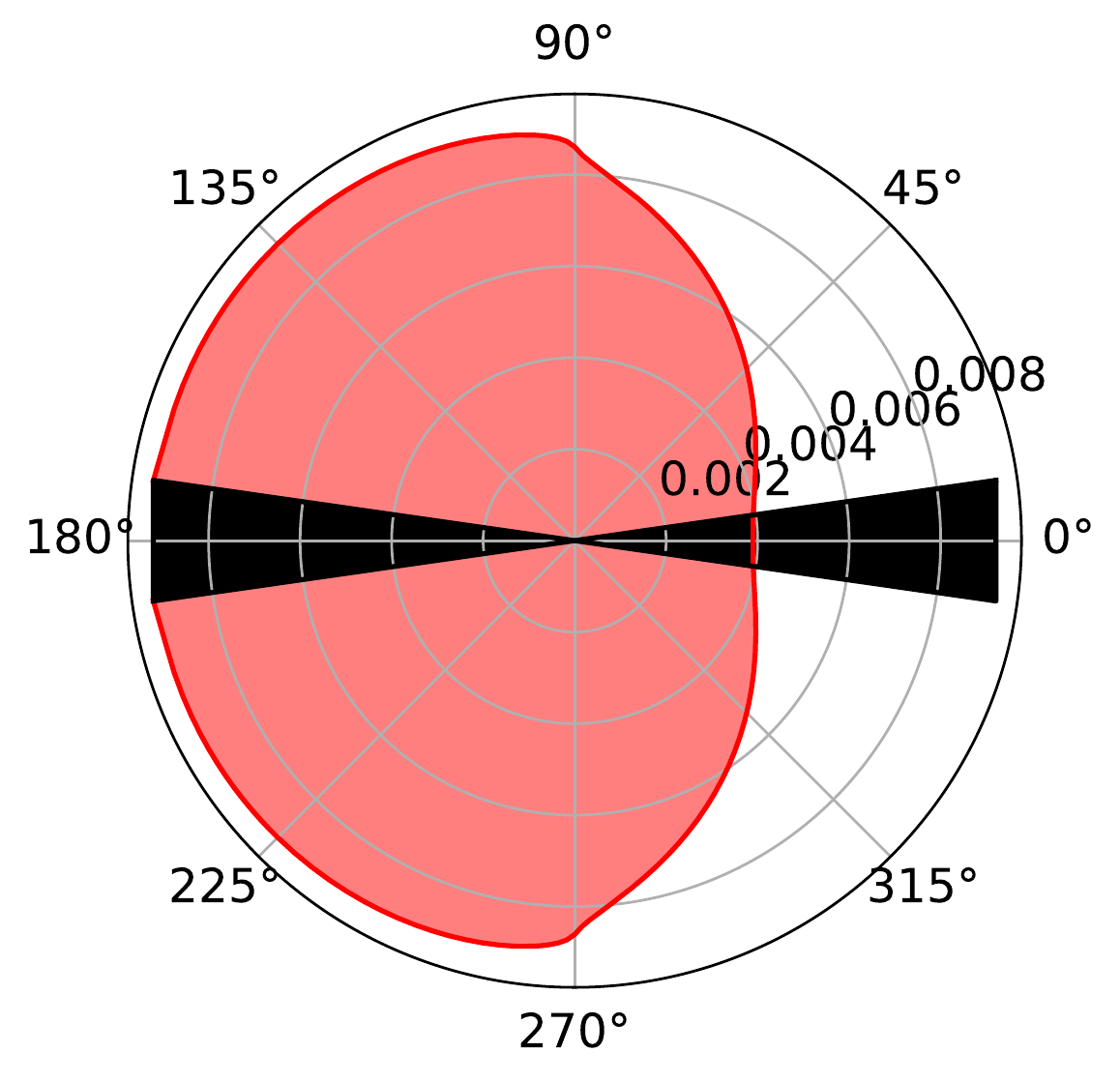}
\includegraphics[width=0.23\textwidth]{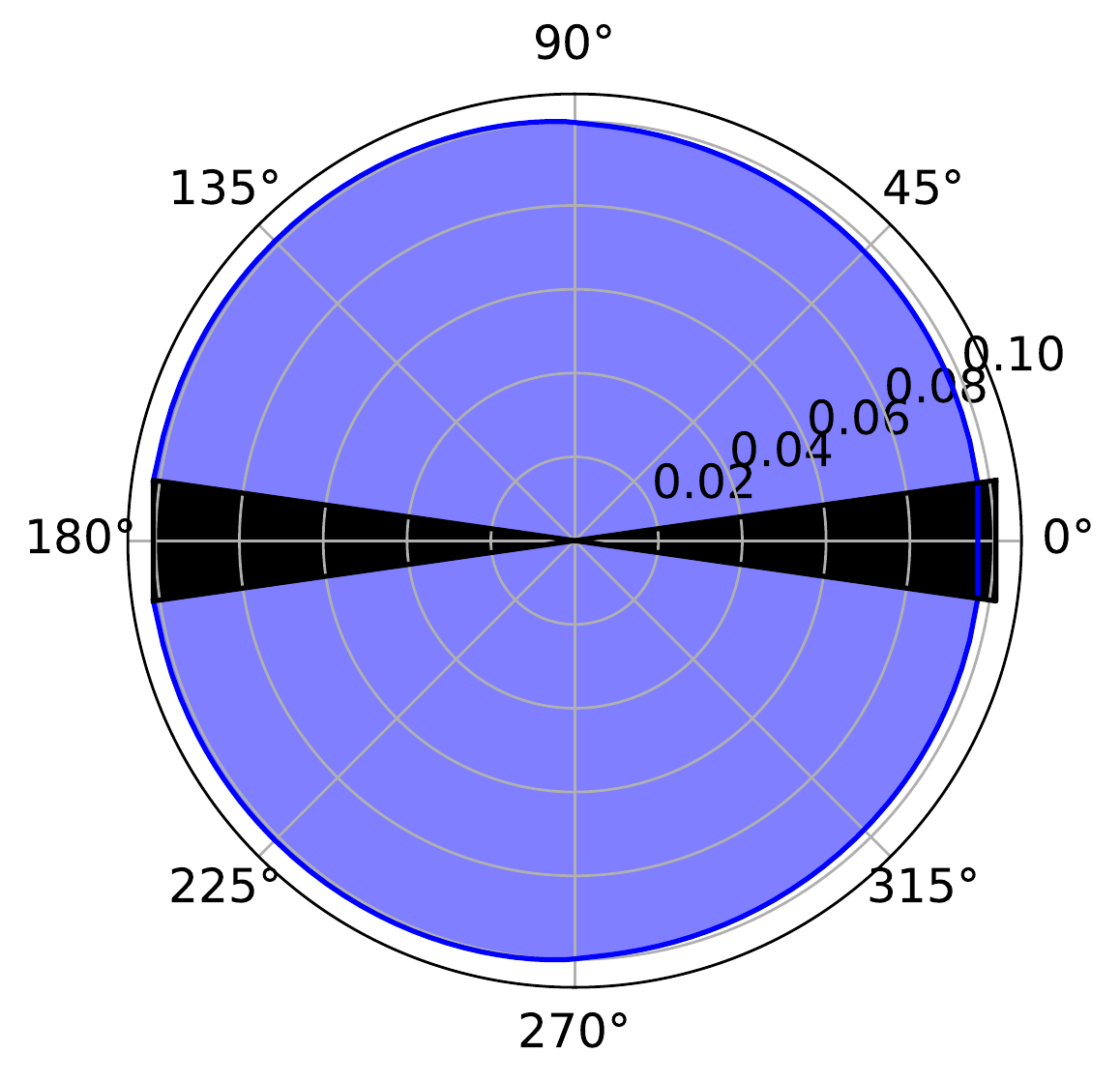}
\includegraphics[width=0.23\textwidth]{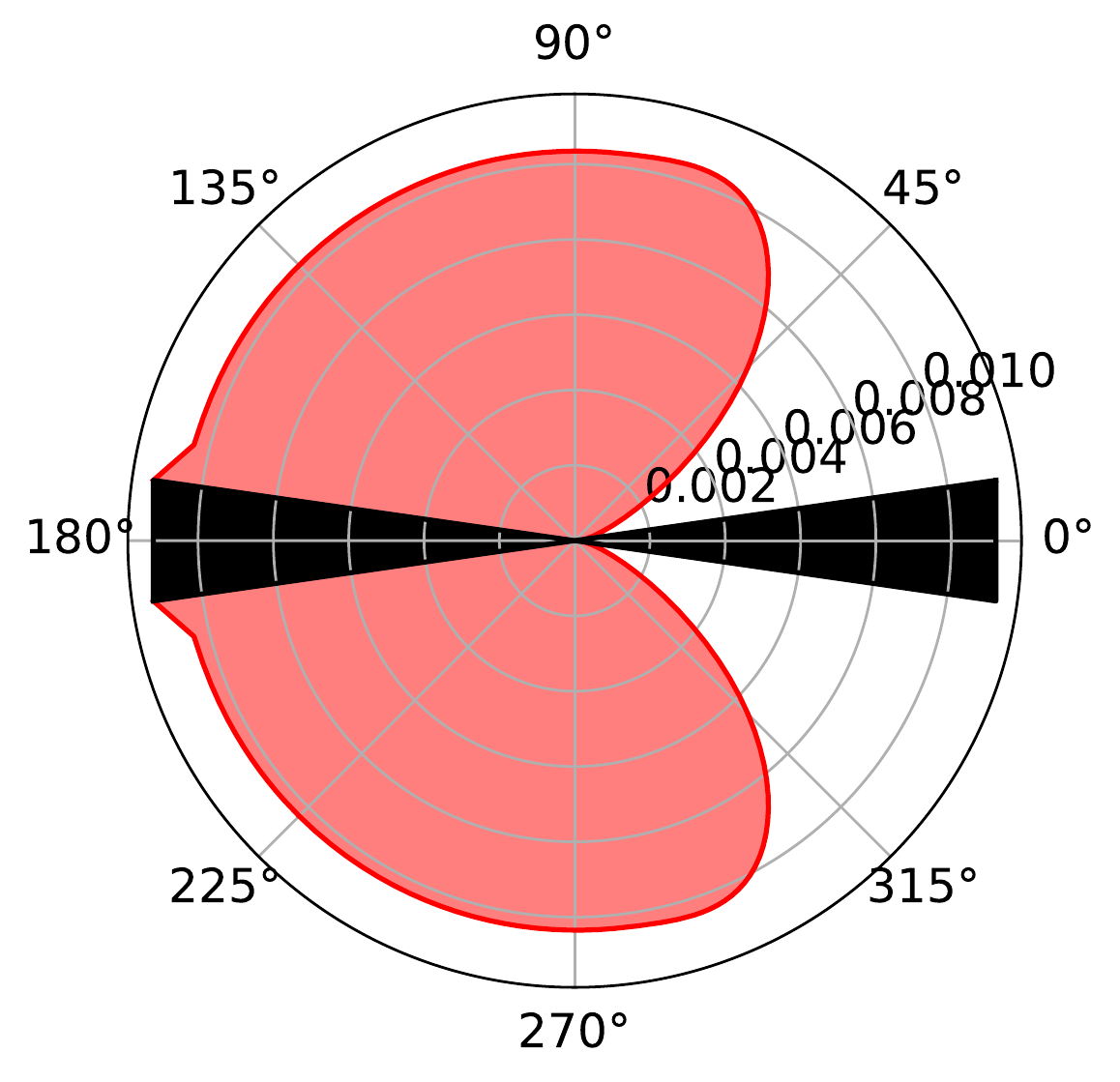}
\includegraphics[width=0.23\textwidth]{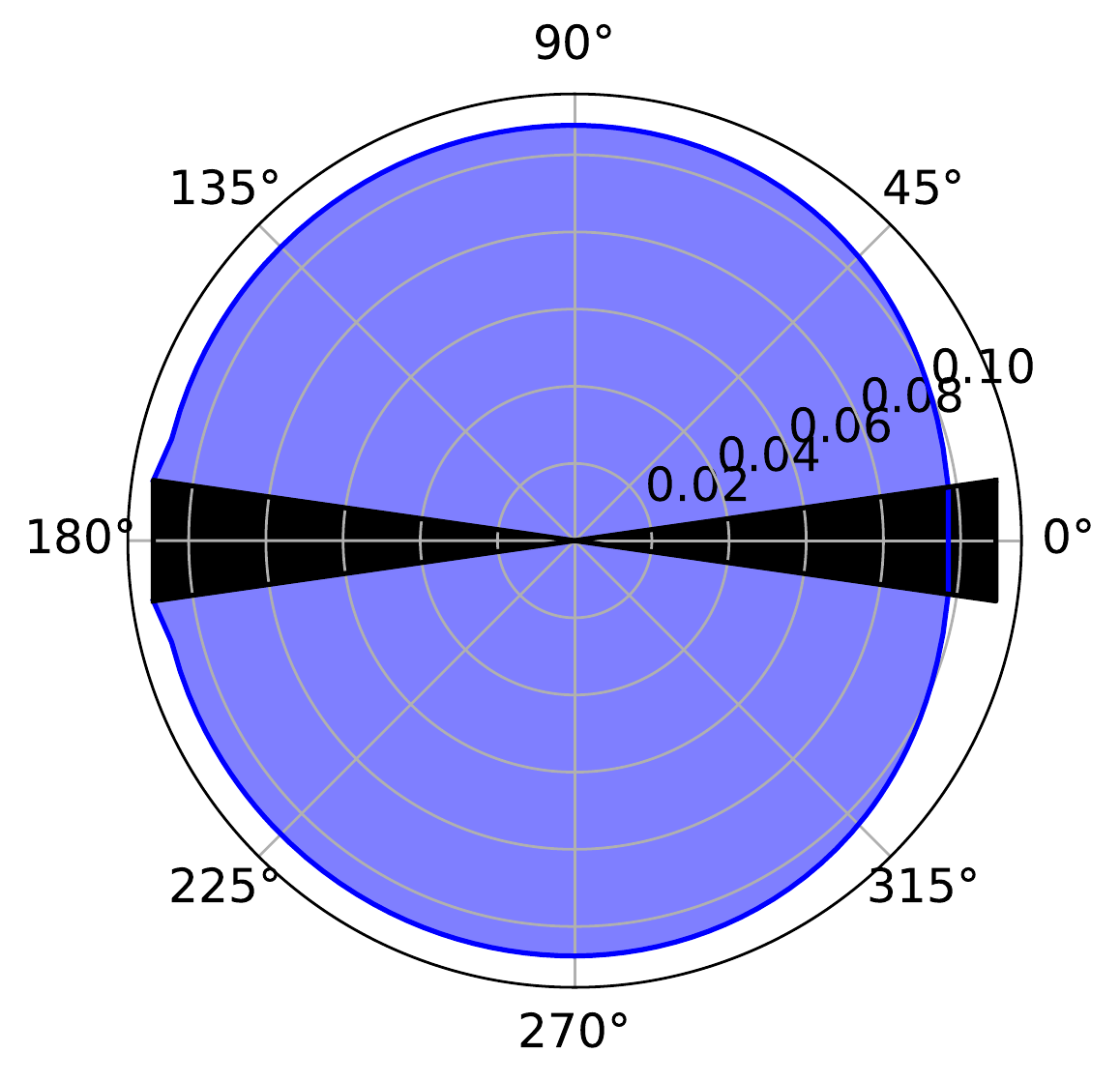}\\
\includegraphics[width=0.23\textwidth]{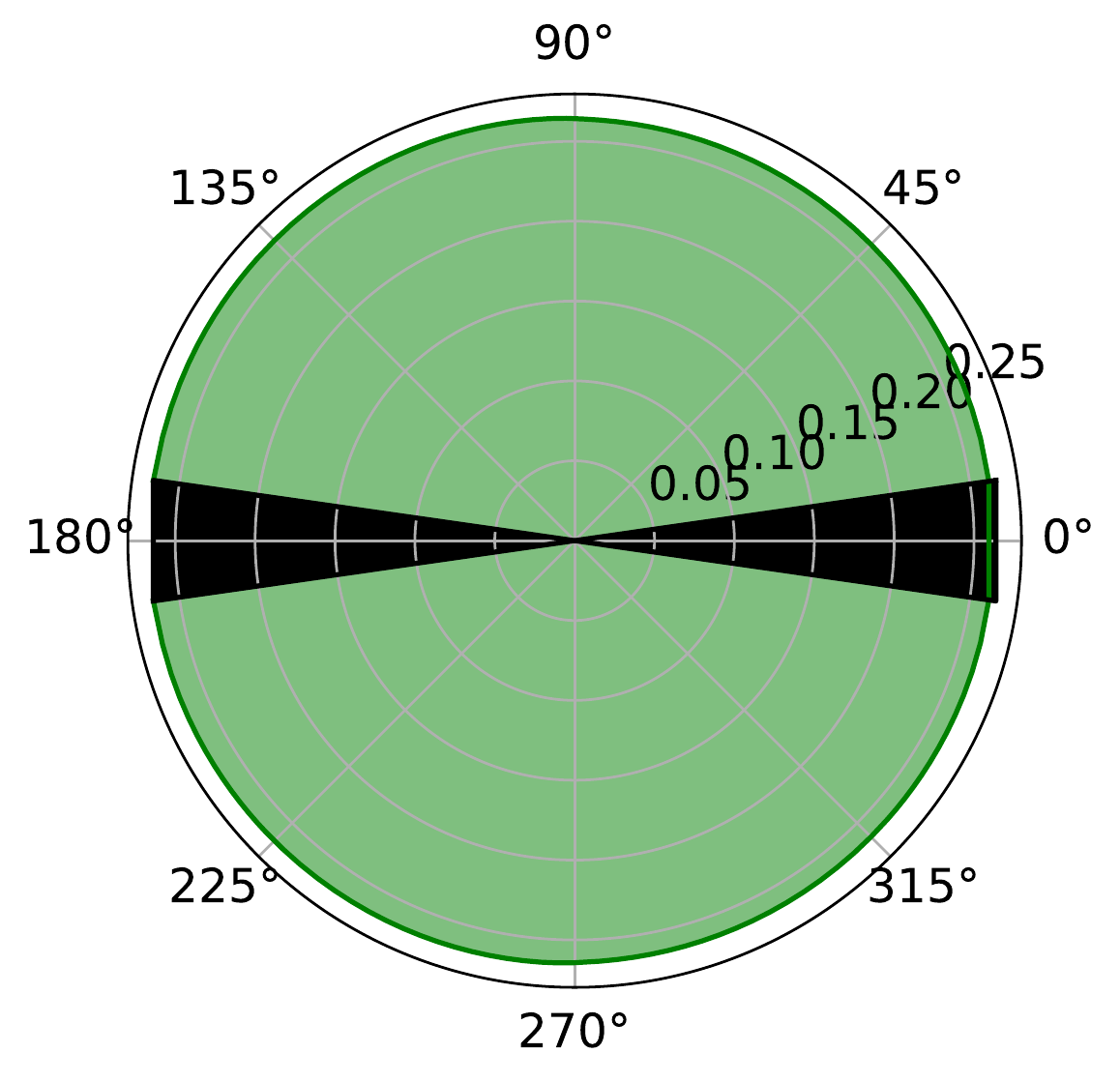}
\includegraphics[width=0.23\textwidth]{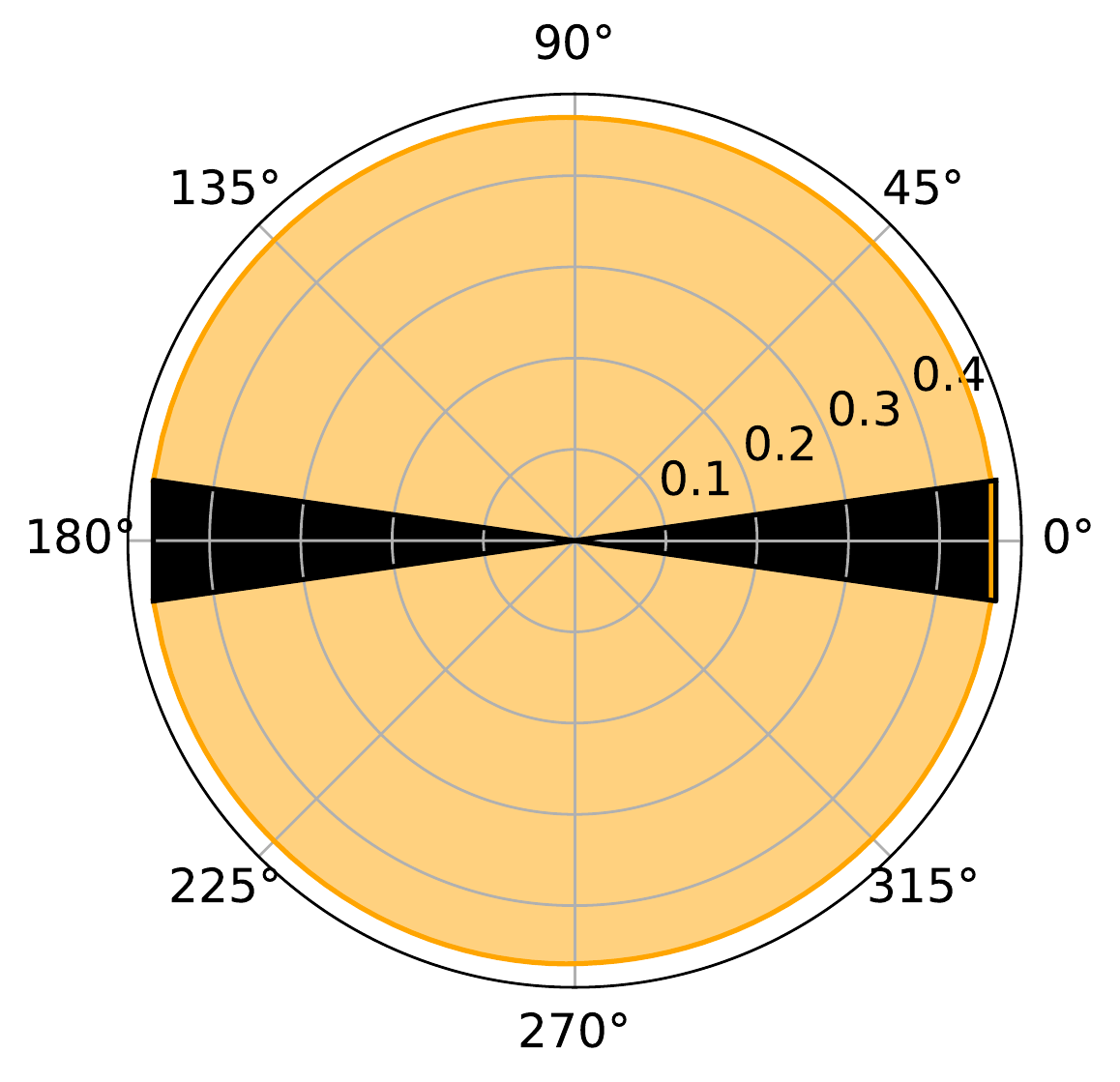}
\includegraphics[width=0.23\textwidth]{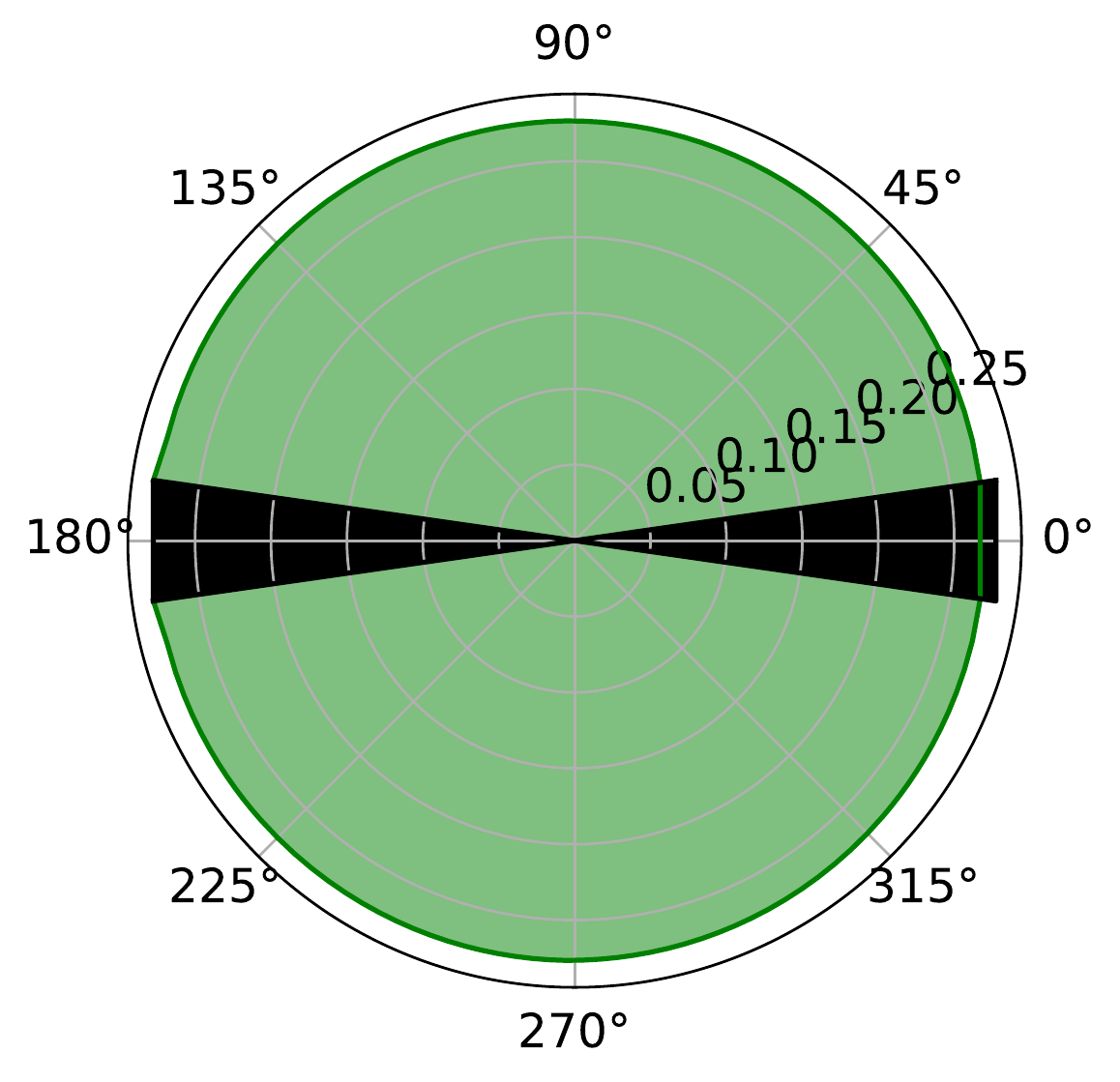}
\includegraphics[width=0.23\textwidth]{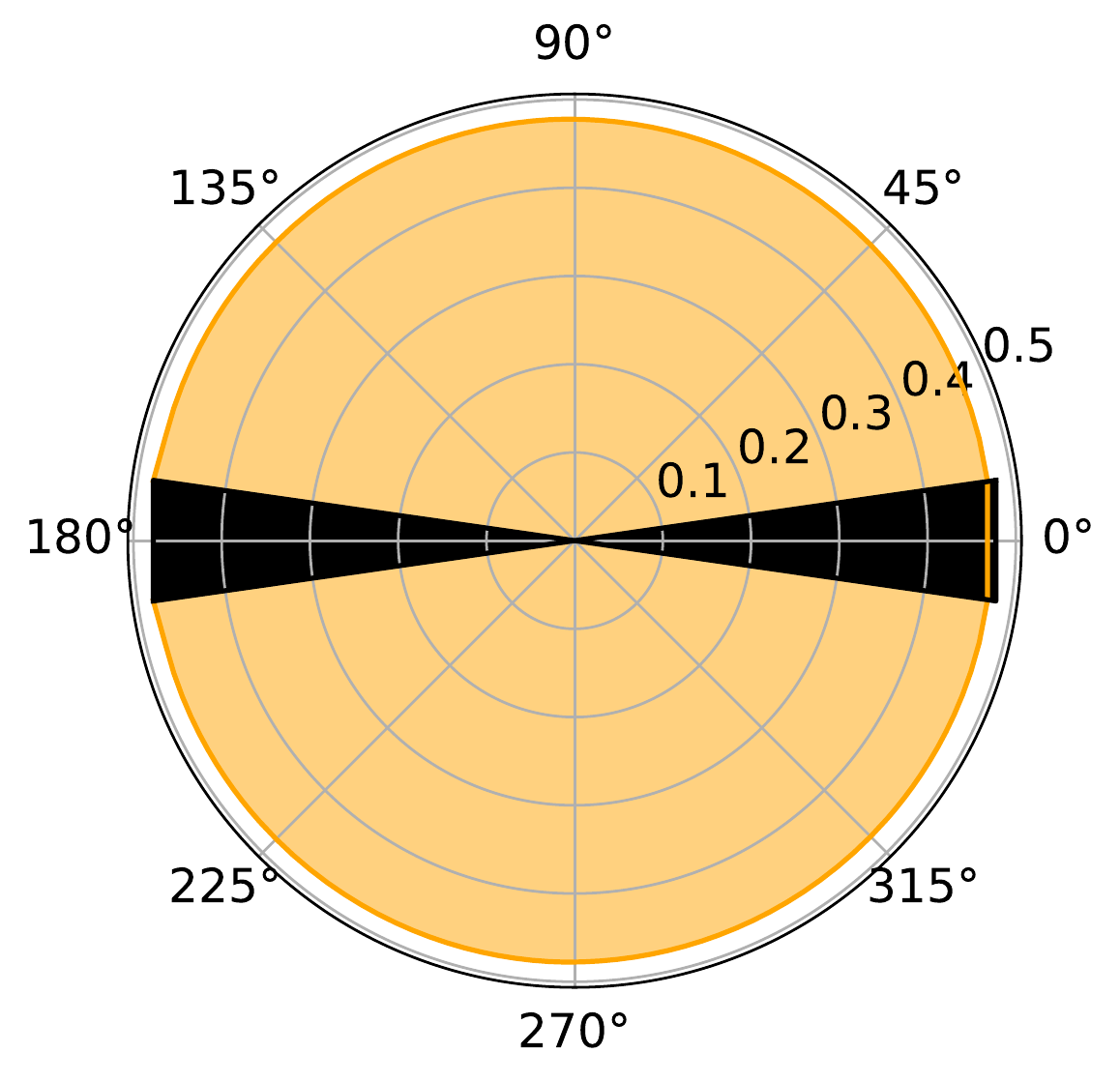}
    \caption{Simulation results and analytical approximations of the intensity, gradient, and anisotropy (top panels) of the GCR distribution as a function of magnetic distance, $z$. The bottom panels show the resulting particle anisotropy as calculated from the numerical model at distances of 0.1 (red), 0.3 (blue), 0.5 (green), and 0.7 (yellow) au. The left panels are for $\lambda = 0.1 \text{ au}$, while the right panels correspond to $\lambda = 1 \text{ au}$. {In the bottom panels, $180^{\circ}$ labels GCRs propagating along the magnetic field towards the Sun and $0^{\circ}$ particles propagating along the magnetic field away from the Sun.}}
    \label{fig:lambdas}
\end{figure*}

\section{Simulation and analytical results}

For the simulation and analytical results presented in this section, we specify an isotropic GCR distribution, with an associated (normalized) omni-directional intensity of $F(1.2\text{ au}) = 100$, at the position of Earth, i.e. $z = 1.2$ au. At the inner boundary, located at $z = 0.05$ au -- the approximate position of the Alfv\'en radius recently crossed by the {\it PSP} spacecraft \citep[][]{kasperetal2021} -- an absorbing boundary condition is assumed where $F(0.05\text{ au}) = 0$. {Unfortunately we cannot extend the simulations below the Alfv\'en radius as the magnetic fields in this region are expected to deviate significant from the large-scale Parker magnetic field assumed in this work. } 

From the omni-directional intensity, we also compute and show the spatial gradient 

\begin{equation}
    	g(z) = \frac{1}{F} \frac{\partial F}{\partial z}
\end{equation}

which is multiplied by $100\%$ in all simulation results to be comparable to measurements which are usually expressed in this fashion. While, usually, a radial gradient is calculated and/or measured, the gradient calculated here is directed along the mean HMF, so that we can refer to it as a {\it magnetic field aligned gradient}.

\subsection{Isotropic formulation with no focusing}

We start with solving Eq. \ref{Eq:diff_eq} by assuming a steady state solution, $\partial F / \partial t = 0$, from which we directly obtain

\begin{equation}
    \frac{\partial F}{\partial z} = \text{constant} \Rightarrow F = \text{constant} \cdot z,
\end{equation}

with a gradient of $g(z) = 1/z$ and, of course, an anisotropy value of $A = 0$.

\subsection{Nearly isotropic formulation with weak focusing}

A steady-state assumption for Eq. \ref{Eq:pseudo_diff_eq} directly leads to a gradient of 

\begin{equation}
    \frac{1}{F} \frac{\partial F}{\partial z} = \frac{1}{L} = g(z),
\end{equation}

while $F(z)$ can be obtained by substituting the definition of $L(z)$, given by Eq. \ref{Eq:def_foc_length}, and solving for $F(z)$ in terms of the magnetic field $B(z)$, to obtain $d \ln F = - d \ln B$, which in turn leads to

\begin{equation}
    F(z) \cdot B(z) = \text{constant}.
\end{equation}



By using Eq. \ref{Eq:streaming_flux} for the streaming flux, the anisotropy follows as

\begin{equation}
    A(z) = \frac{3S}{vF} = - \frac{\lambda_{||}}{L(z)}.
\end{equation}

Note that this expression diverges for strong {focusing/mirroring}, $A \rightarrow - \infty$ as $L \rightarrow 0$. This is due to the assumption of weak {focusing/mirroring} used when driving Eq. \ref{Eq:pseudo_diff_eq}, i.e. $\lambda_{||} \ll L$. This strict condition is not satisfied for all parameters assumed in this work.
\subsection{Focused transport equation}

For the simulations presented here we assume a parametrized version of $D_{\mu\mu}$ from quasi-linear theory \citep[][]{jokipii1966}, 

\begin{equation}
    D_{\mu \mu} = D_0 (1 - \mu^2)\left( |\mu|^{q - 1} + H \right),
\end{equation}

with $q = 5/3$ representing the inertial range turbulence spectral index and $H=0.05$ accounting for possible dynamic effects in an {\it ad-hoc} fashion \citep[e.g.][]{beeckwibberenz1986}. We do not, however, calculate $D_0$ from first principles \citep[as is done by e.g.][]{straussetal2017a}, but rather specify $\lambda_{||}$ from which $D_0$ can be calculated from Eq. \ref{Eq:lamda_def}. Eq. \ref{Eq:focused_eq} is solved with the open-source numerical model presented in \citet{vandenBergetal2020} and available from \url{https://github.com/RDStrauss/SEP_propagator}. 

The resulting particle intensities, for different choices of $\lambda_{||}$ are shown in Fig. \ref{fig:lambdas}. For all simulations, protons with a kinetic energy of 100 MeV is assumed. However, with this specific model set-up (including an energy independent $\lambda_{||}$, no energy losses, and the assumption of a steady-state solution) the results are independent of the particle energy considered.

The left panels of Fig. \ref{fig:lambdas} corresponds to the case of $\lambda_{||} = 0.1 \text{ au}$, while the right panels correspond to $\lambda_{||} = 1 \text{ au}$. {Although the mean-free-path is treated, in this work, as a free parameter, it is chosen to correspond to consensus values at Earth from e.g. \citet{bieberetal1994}, who reported values ranging from $\sim 0.1 - 1$ au for protons in the rigidity range of $\sim 10 - 10^4$ MV.} The top panels show the computed omni-directional intensity, the calculated gradient, and the resulting anisotropy as a function of magnetic distance, $z$. Results from the numerical model, solving Eq. \ref{Eq:focused_eq}, are shown as red symbols, the case of weak scattering, i.e. solving Eq. \ref{Eq:pseudo_diff_eq}, as the solid blues line, and the case of the isotropic diffusion equation, Eq. \ref{Eq:diff_eq} as the gray dashed lines. The intensities, and hence also the gradients, as calculated by the numerical model and the analytical approach assuming weak {focusing/mirroring} compares relatively well. However, the analytical approach cannot capture the radial dependence of the anisotropy as $\lambda_{||} \ll L$ is generally violated for all assumed parameters. The diffusion equation, assuming no {focusing/mirroring}, gives a smaller gradient as compared to the other approaches.

The bottom panels of Fig. \ref{fig:lambdas} show the pitch-angle distribution of the particles at different radial positions (these positions are indicated by the vertical arrows in the upper panels) as calculated form the numerical model. In these graphs, $\alpha (\mu) = 0^{\circ}(1)$ indicates particles streaming away from the Sun and $\alpha (\mu) = 180^{\circ} (-1)$ labels particles moving towards the Sun. The black shaded regions show the pitch-angle regions close to the model's boundaries where information about the the particle distribution is not available. For the assumed values of $\lambda_{||}$, strong anisotropies, and significant anisotropic behaviour, is only observed very close to the inner boundary (e.g. inside 0.1 au). 

\section{Discussion}

{While the loss-cone distributions associated with the \citet{parker1958} HMF predict highly anisotropic GCR distributions in the inner heliosphere (see Fig. \ref{fig:loss_cone}), the inclusion of scattering tends to lead to more isotropic distributions (see Fig. \ref{fig:lambdas}), except in the very inner heliosphere, i.e. $z < 0.2$ au. In general the cosmic ray distribution will be isotropic if pitch-angle scattering dominates the {focusing/mirroring} process, i.e. when $\lambda_{||} \ll L$. While $L$ is determined by the \cite{parker1958} HMF geometry, $\lambda_{||}$ depends (in a non-trivial fashion) on the underlying turbulence structure and how particles resonate with these fluctuations, making $\lambda_{||}$ difficult to quantify, especially in the newly-explored regions closer to the Sun. However, even when adopting simplified forms of $\lambda_{||}$, results of our numerical model suggest small but measurable first-order anisotropies in this region, increasing in magnitude towards the Sun, with a deficiency of outwards (i.e. away the Sun) propagating GCRs due to the mirroring effect. Larger anisotropies are expected for particles with larger values of $\lambda_{||}$, such as high-rigidity cosmic rays during solar minimum conditions. }\\

Interestingly, our results generally show that by assuming an isotropic distribution governed by a diffusion equation, and therefore neglecting particle {focusing/mirroring} near the Sun, leads to an under-estimation of the spatial gradient when compared to the numerical solution. Potentially this can explain the discrepancy between measured and calculated gradients in the very inner heliosphere as shown by \citet{rankinetal2021,rankinetal2021b} for anomalous cosmic rays. Additionally, this suggests that traditional modulation models, based on solutions of the \cite{parker1965} transport equation, are not valid close to the Sun where focusing/mirroring forces have appreciable effects.\\

{This study focused on particle transport along the mean magnetic field (essentially a 1D spatial geometry assuming scattering by quasi-wavelike parallel-propagating turbulence) and neglected any cross-field transport processes including drifts and perpendicular diffusion that will lead to 2D transport effects. At present it is not clear whether such cross-field processes will have appreciable effects on particle anisotropies, especially given recent {\it PSP} measurements of a dominating quasi-parallel turbulent component inside 1 au \citep[][]{BandyopadhyayMcComas2021,zhaoetal2022} and the fact that the higher turbulence levels in this region will significantly disrupt the drift process \citep[][]{vandenBergetal2021}. Additional, it should be kept in mind that pitch-angle scattering is primarily responsible for the isotropization of the distribution function, and given the fact that we do include this process in our modelling approach, we do not expect the results presented here to qualitatively change when additional cross-field processes are included in future work. {Additionally, \citet{ruffolo1995} showed that the effect of adiabatic energy losses becomes negligible for solar energetic particle protons above 20 MeV. As such, we do not expect that the inclusion of adiabatic energy losses will have any appreciable effects on the simulation results presented here for higher energy GCRs.}}

\section{Conclusion}

Traditional cosmic ray transport models neglect the magnetic mirroring effect close to the Sun. However, with the {\it PSP} spacecraft currently exploring the very inner heliosphere, it has become necessary to quantify possible mirroring effects on the ACR and GCR distributions. Such a study is presented in this work where we have included the effects of both magnetic mirroring and pitch-angle scattering. Our results predict measurable cosmic ray anisotropies close to the Sun, i.e. inside $z < 0.2$ au, that generally increases with increasing particle rigidity (i.e. increasing $\lambda_{||}$) towards the Sun. 

\acknowledgments

This work is based on the research supported in part by the National Research Foundation of South Africa (NRF grant numbers 119424, 120345, and 120847). Opinions expressed and conclusions arrived at are those of the authors and are not necessarily to be attributed to the NRF. The responsibility of the contents of this work is with the authors. Figures prepared with Matplotlib \citep{hunter} and certain calculations done with NumPy \citep{harrisetal2020}.



\end{document}